\documentstyle[twoside,fleqn,espcrc2]{article}

\newcommand{\AmS}{{\protect\the\textfont2
  A\kern-.1667em\lower.5ex\hbox{M}\kern-.125emS}}
\global\arraycolsep=2pt
\title{Finite-element quantum field theory\hfill OKHEP-96-06}

\author{Kimball A. Milton\address{Department of Physics and
Astronomy,
         The University of Oklahoma\\
        Norman, OK 73019-0225 USA}%
        \thanks{E-mail: kmilton@ou.edu.  Supported in part by the
                U.S. Department of Energy.}}

\begin{document}

\begin{abstract}
An alternative approach to lattice gauge theory has been under
development
for the past decade.  It is based on discretizing the operator
Heisenberg
equations of motion in such a way as to preserve the canonical
commutation
relations at each lattice site.  It is now known how to formulate a
non-Abelian
gauge theory within this framework.  The formulation appears to be
free
of fermion doubling.  Since the theory is unitary, a time-development
operator (Hamiltonian) can be constructed.
\end{abstract}

\maketitle

\section{QUANTUM MECHANICS}
Let us consider a one-dimensional field theory, namely quantum
mechanics,
governed by a continuum Hamiltonian $H=p^2/2+V(q)$.  The
corresponding
Heisenberg equations are
\begin{equation}
\dot p=-V'(q),\quad \dot q=p.
\label{heisen}
\end{equation}
These equations are to be solved subject to the initial condition
$[q(0),p(0)]=i$.  The equations (\ref{heisen}) preserve unitarity in
the sense that at any subsequent time $t$ the canonical commutation
relations
continue to hold, $[q(t),p(t)]=i$.
\subsection{Finite-element discretization}
We discretize the above equations of motion by dividing the time
interval
$[0,T]$ into $N$ equal intervals of length $h$.  On each interval we
define a local time variable $t$, $0\le t\le h$, and write the
dynamical
variables as $r$th degree polynomials in $t$,
\begin{equation}
p(t)\approx\sum_{k=0}^r a_k(t/h)^k,\quad q(t)\approx \sum_{k=0}^r
b_k(t/h)^k.
\end{equation}
We determine the $2r+2$ operator coefficients, $a_k$, $b_k$, by
imposing
continuity between the intervals, that is, at the lattice sites,
and by imposing the equations (\ref{heisen})
at $r$ different points in the interval, at $t_i=\alpha_i h$.
Unitarity,
that is, that $[q_n,p_n]=i$ {\em exactly\/} at each lattice site,
then uniquely determines that the points at which the equations
of motion are to be imposed are the Gaussian knots, the solutions
of $P_r(2\alpha-1)=0$.  Remarkably, these are precisely the points
at which classically the numerical error is minimized, the relative
error  after $N$ steps then going like $N^{-2r}$.

\subsection{Time-evolution operator}

Because the lattice theory is unitary, there must exist a unitary
time-evolution
operator which advances operators through the lattice,
\begin{equation}
q_{n+1}=Uq_nU^{-1},\quad p_{n+1}=U p_nU^{-1}.
\end{equation}
For linear finite elements, $r=1$, $U$ is given by
\begin{eqnarray}
U&=&e^{ihp_n^2/4}e^{ihA(q_n)}e^{ihp_n^2/4},\\
A(q)&=&{2\over h^2}[q-g^{-1}(4q/h^2)]^2+V[g^{-1}(4q/h^2)],\\
g(x)&=&{4\over h^2}x+V'(x).
\end{eqnarray}
It is apparent that this is an implicit scheme, as the equations of
motion must be solved to construct the operator $A$.  Of course, we
can use this evolution operator to define a self-adjoint lattice
Hamiltonian, ${\cal H}=-{i\over h}\ln U$, which differs from the
continuum Hamiltonian by $O(h^2)$.  [The order $r$ finite element
prescription would give a Hamiltonian differing from the continuum
one by $O(h^{2r})$].  The details are given in \cite{review}.

\subsection{Harmonic-Oscillator Matrix Elements of the Time-Evolution
Operator}

Remarkably, in view of the above, a closed expression, involving
$g$ not $g^{-1}$, can be derived for matrix elements of the transfer
operator in a harmonic oscillator basis \cite{das,bss}.  Although
this
formula is fairly complex,  it can be easily used to extract
expansions
in powers of the lattice spacing $h$.  For the anharmonic oscillator
with potential $V=\lambda q^{2k}/2k$, we find in the harmonic
oscillator
groundstate
\begin{equation}
\langle
U\rangle=1+ih\lambda^{1/(k+1)}f(\alpha)-{h^2\over2}\lambda^{2/(k+1)}
s(\alpha),
\end{equation}
where $\alpha=\lambda\gamma^{2k+2}$, $\gamma$ is the width of the
harmonic oscillator state, and
\begin{eqnarray}
f(\alpha)&=&{1\over4\alpha^{1/(k+1)}}\left(1+{2\alpha\over
k}{\Gamma(k+1/2)
\over\Gamma(1/2)}\right)\\
s(\alpha)&=&{1\over16\alpha^{2/(k+1)}}\left(3-4\alpha{2k-1\over k}
{\Gamma(k+1/2)\over\Gamma(1/2)}\right.\nonumber\\
&&\mbox{}+\left.{4\alpha^2\over k^2}{\Gamma(2k+1/2)\over
\Gamma(1/2)}\right).
\end{eqnarray}
We match this expansion to that of $\exp(i\omega_0 h)$ to determine
approximations to the ground state energy.  If we use the order $h$
data
only, we must determine the variational parameter $\alpha$ by,
say, the principle of minimum sensitivity.  The energy so determined
is
accurate to a few percent.  If the order $h^2$ data is used $\alpha$
is
determined, and the results are considerably
more accurate.  If, instead, we use a
two-state approximation to the anharmonic oscillator ground state,
improvement by more than an order of magnitude results.
See Table \ref{tab1}.  Wavefunctions are equally impressive.

\begin{table*}[hbt]
\setlength{\tabcolsep}{1.4pc}
\newlength{\digitwidth} \settowidth{\digitwidth}{\rm 0}
\catcode`?=\active \def?{\kern\digitwidth}
\caption{Ground-state energies for a $\lambda q^2k/2k$ potential.
(Figures in parentheses are percentage relative errors.)}
\label{tab1}
\begin{tabular*}{\textwidth}{llllll}
\hline
             $k$    & \multicolumn{2}{c}{$O(h)$}
                 & \multicolumn{2}{c}{$O(h^2)$}&Exact \\
\cline{2-3} \cline{4-5}
                 & \multicolumn{1}{c}{One state}
                 & \multicolumn{1}{c}{Two state}
                 & \multicolumn{1}{c}{One state}
                 & \multicolumn{1}{c}{Two state}         \\
\hline
2&$0.4293(+2)$ & $0.4212(+0.1)$ & $0.4178(-0.7)$ &
$0.4205(-0.06)$&$0.42081$ \\
3 &$0.4639(+7)$ & $0.4391(+0.9)$ & $0.4453(+2)$&
$0.4328(-0.5)$&$0.43493$ \\
4& $0.5230(+13)$ & $0.4772(+3)$ & $0.5171(+11)$ &
$0.4647(-0.2)$&$0.46450$ \\
\hline
\end{tabular*}
\end{table*}

\section{NON-ABELIAN GAUGE THEORY}

Recently it was discovered how to implement non-Abelian gauge
invariance
on a linear finite-element lattice \cite{nagt}.  This was done
constructively,
by gauging the free lattice Dirac and Yang-Mills equations (which, by
the
linear finite-element prescription, are forward difference equations,
forward-averaged in the non-differentiated directions), and
simultaneously
determining the form of the gauge transformation.  The result is most
conveniently expressed in terms of the following {\em link operator}
defined on a particular finite-element (hypercube):
\begin{equation}
(L_\mu)_{\bf m}=\exp(-igh({\cal A}_\mu)_{\bf m}),
\end{equation}
where $\bf m$ is a local four-vector coordinate, having components 0
or 1, indicating the 16 corners of the hypercube, and, in terms of
absolute coordinates
\begin{equation}
({\cal A}_\mu)_{m_\mu,m_\perp}=(A_\mu)_{\overline{m_\mu-1},m_\perp},
\end{equation}
with the overbar indicating forward averaging in the indicated
coordinate,
and $\perp$ indicating coordinates other than the one specified.
The result of the above construction is that the link operator
transforms
very simply,
\begin{equation}
\delta(L_\mu)_1=ig[\delta\omega_0(L_\mu)_1-(L_\mu)_1\delta\omega_1],
\end{equation}
where only the local value of the $m_\mu$ coordinate is shown (the
other
coordinates are the same throughout).
The covariant, ``transversely local'' field strength is
a path ordered product of link operators around the $\mu$-$\nu$
plaquette:
\begin{eqnarray}
&&\exp(-igh^2({\cal F}_{\mu\nu})_m)=P\exp(-ig\oint A\cdot
dl)\nonumber\\
&&\qquad=(L_\mu)_{10}(L_\nu)_{11}(L^\dagger_\mu)_{11}
(L^\dagger_\nu)_{01},
\end{eqnarray}
where now the local coordinates refer to the values of $m_\mu$ and
$m_\nu$.
This, however, does not reduce to the free finite element form if
$g=0$,
and therefore would violate unitarity.  So we introduce a covariant
averaging operator,
\begin{equation}
(\tilde{\cal D}_\lambda
X)_0={1\over2}[(L_\lambda)_1X_1(L^\dagger_\lambda)_1
+X_0],
\end{equation}
where now the coordinate refers to $m_\lambda$.  The field strength
is
then (symmetric averaging)
\begin{equation}
(F_{\mu\nu})_{\overline{m}}=\left(\prod_{\lambda\ne\mu,\nu}
\tilde{\cal D}_\lambda{\cal F}_{\mu\nu}\right)_m.
\end{equation}
The field strength transforms covariantly in the sense
\begin{equation}
\delta(F_{\mu\nu})_{\overline{m}}=ig[\delta\omega_m,
(F_{\mu\nu})_{\overline{m}}].
\label{gt}
\end{equation}

\subsection{Yang-Mills equations}

Although the above field strength construction is {\em local}, it is
not possible to obtain local Dirac and Yang-Mills equations without
sacrificing unitarity \cite{quasilocal}.  The inductive approach
in \cite{nagt} gives the following covariant Yang-Mills equations
\begin{equation}
\sum_{\nu\ne\mu}\left\{{1\over h}[(F^{\mu\nu})_1
-(F^{\mu\nu})_0]
+{\cal K}^\nu_{0}\right\}=j^\mu_{0},
\label{ym}
\end{equation}
where the value of the local $\nu$ coordinate is shown (the curl
involves averaging in $m_\perp$), and the
interaction term ${\cal K}^\nu_n={\cal K}_n$
can be easily constructed from the difference equation
\begin{equation}
{\cal K}_{n-1}L_n+L_n{\cal K}_{n}=
{2\over h}[L_n,(F^{\mu\nu})_{n,\overline{m}_\perp}],
\end{equation} where $L_n=L^{\nu}_n$.  The current in (\ref{ym})
may be taken to be the local form,
\begin{equation}
j^\mu_m=g\overline\psi_{\overline{m}}T\gamma^\mu\psi_{\overline{m}},
\label{current}
\end{equation}
which transforms like (\ref{gt}).
\section{AXIAL-VECTOR ANOMALIES AND FERMION DOUBLING}

For an Abelian theory it is easy to obtain an explicit form for the
transfer matrix, which carries the Dirac operator from one time to
the
next, $\psi_{n+1}=T_n\psi_n$.  In the temporal gauge, $A_0=0$:
\begin{equation}
T={1-ih\mu\gamma^0/2+\gamma^0\vec\gamma\cdot\vec{\cal D}\over
1+ih\mu\gamma^0/2-\gamma^0\vec\gamma\cdot\vec{\cal D}},
\end{equation}
where $\vec{\cal D}$ is a covariant derivative operator
\cite{absence}.
Using this, nonzero vector and axial-vector anomalies have been
computed
in 2 and 4 space-time dimensions \cite{absence,quasilocal}.  For
example,
in 2 dimensions
\begin{equation}
\langle\mbox{``}\partial_\mu J_5^\mu\mbox{''}\rangle
={e^2E\over M\sin\pi/M}\left\{\begin{array}{c}
\cos^2\pi/2M\\1\end{array}\right\},
\end{equation}
for $M$ odd or even, respectively,
where $E$ is the electric field and $M$ is the number of spatial
lattice
sites.  Note this reduces to continuum form, with error, as expected,
of
$O(M^{-2})$, as $M\to\infty$. At the same time we can compute the
lattice
vacuum polarization (one fermion loop) and obtain results very close
to  the expected value
of $e^2/\pi$.  So there is no evidence of species doubling, which
freedom is
already signaled by the form of the free dispersion relation
\cite{bms},
\begin{equation}
\omega^2=\sum_{i=1}^3{4\over h^2}\tan^2{p_i\pi\over M}+\mu^2.
\end{equation}

Note that the Minkowski-space formulation is crucial to this
conclusion.
If we were working in Euclidean space, it is possible to define an
action
from which the equations of motion could be derived.  Since that
action possess
chiral symmetry, the currents derived from it (which involve all
times,
``past'' and ``future'') will be anomaly free.
Explicitly, fermion doublers then appear and contribute to the vacuum
polarization \cite{quasilocal}.

But in Minkowski space, it is not possible to derive an action.
Therefore,
there is no connection between symmetry and conserved currents.  In
fact,
the current (\ref{current}) must be freely invented (consistent with
gauge
invariance), and can, and does, possess anomalies.  Explicitly, the
vacuum
polarization gets its sole contribution from the neighborhood of
zero-momentum
fermions.


\begin{thebibliography}{9}
\bibitem{review} C. M. Bender, L. R. Mead, and K. A. Milton,
Computers Math. Applic. {\bf 28} (1994) 279.
\bibitem{das} K. A. Milton and R. Das, Lett. Math. Phys. {\bf 36}
(1996) 177.
\bibitem{bss} C. M. Bender, L. M. Simmons, Jr., and R. Stong, Phys.
Rev.
D {\bf 32} (1986) 2362.
\bibitem{nagt} K. A. Milton, Nucl. Phys. {\bf B452} (1995) 401.
\bibitem{quasilocal} K. A. Milton, Phys. Rev. D {\bf 53} (1996) 5898.
\bibitem{absence} K. A. Milton, Lett. Math. Phys. {\bf 34} (1995)
285.
\bibitem{bms} C. M. Bender, K. A. Milton, and D. H. Sharp, Phys. Rev.
Lett.
{\bf 51} (1983) 1815.
\end{thebibliography}
\end{document}